\documentclass[a4paper,fleqn,usenatbib]{mnras}

\usepackage{newtxtext,newtxmath}

\usepackage[T1]{fontenc}
\usepackage{ae,aecompl}

\usepackage{graphicx}	
\usepackage{amsmath}	
\usepackage{xfrac}		

\newcommand{\s}[1]{\textrm{#1}}

\title[Hunt for dark subhaloes]{Hunt for dark subhaloes in the galactic stellar field using computer vision}

\author[Mihael Peta\v c]{
Mihael Peta\v c,$^{1,2}$\thanks{E-mail: petac@lupm.in2p3.fr}
\\
$^{1}$SISSA \& INFN -- Sezione di Trieste, Via Bonomea 265, 34136 Trieste, Italy\\
$^{2}$Laboratoire Univers et Particules de Montpellier (LUPM) \& CNRS \& Universit\'e de Montpellier (UMR-5299), \\ Place Eug\`ene Bataillon, F-34095 Montpellier Cedex 05, France}

\date{Accepted XXX. Received YYY; in original form ZZZ}

\pubyear{2020}

\begin{document}
\label{firstpage}
\pagerange{\pageref{firstpage}--\pageref{lastpage}}
\maketitle

\begin{abstract}
The lack of tangible evidence for non-gravitational interactions between dark and visible sectors drives the need for exploring new avenues of inferring dark matter properties through purely gravitational probes. In particular, addressing small-scale distribution of dark matter could lead to valuable new insights into its particle nature, either confirming predictions of the collisionless cold dark matter hypothesis or favouring models with suppressed small-scale matter power spectrum. In this work we propose a novel machine learning approach for constraining the abundance of galactic dark matter subhaloes through the analysis of Milky Way's stellar field that has been only recently mapped with sufficient coverage thanks to the Gaia mission. Our method is based on convolutional neural networks which represent a powerful tool for identifying characteristic perturbations in spatial maps of stellar number density and velocity distribution moments. For generating the training data we develop a robust and computationally efficient algorithm, capable of generating mock stellar fields from an arbitrary underlying phase-space distribution of stars. By preforming a preliminary study of the outlined approach on synthetic datasets we demonstrate that sensitivities down to (or even below) $10^8 M_\odot$ could be reached. Furthermore, our results show that the accuracy of the advocated technique crucially depends on the  kinematic properties of mapped stars and could be further improved by applying it to abundant stellar populations with particularly low velocity dispersion, such as the galactic thin disc stars.
\end{abstract}




\section{Introduction}
\label{sec:introduction}

One of the characteristic predictions of the cold dark matter (DM) paradigm is hierarchical growth of structure~\cite{white_core_1978,peebles_large-scale_1980,davis_evolution_1985,quinn_numerical_1986}. According to it, the smallest objects collapsed under self-gravity first and subsequently underwent merging which led to the formation of increasingly large structures. This is supported by various observations, spanning from cosmological scales down to a fraction of a Mpc, giving an excellent match to the matter power spectrum predicted within the standard cosmological model~\cite{collaboration_planck_2018,chabanier_matter_2019}. However, the study of DM distribution on sub-galactic scales turns out to be much more demanding since it requires extremely precise astronomical measurements as well as accounting for non-linear gravitational dynamics and baryonic physics. None-the-less, constraining the clustering properties of DM on small scales seems a promising way of obtaining new insights into the nature of DM particles, as many alternatives to the collisionless cold DM hypothesis involve suppression of the matter power spectrum below the scale of to the smallest observed galaxies. Some of the most prominent suggestions of this type are warm DM~\cite{dodelson_sterile_1994,bode_halo_2001,hooper_mev_2007}, self-interacting DM~\cite{spergel_observational_2000,tulin_dark_2018} or ultra-light bosonic DM~\cite{hu_fuzzy_2000,hui_ultralight_2017}. The key benefit of studying the phenomenon of DM through the small-scale power spectrum is the fact that its signatures can be searched for through purely gravitational probes, which have been so far crucial for establishing the known properties of DM. On the other hand, no tangible detection of DM has ever been achieved through non-gravitational probes, such as direct and indirect searches or particle colliders, despite decades of efforts. Furthermore, it is important to note that a better understanding of sub-galactic DM distribution could also help in improving the constraints on DM coupling with visible matter; this could be achieved either through better modelling of local DM distribution which crucially enters the interpretation of direct detection experiments
or improving our understanding of its galactic distribution and discovering new prime targets for indirect searches.

Significant progress in addressing the DM clustering properties on small scales has been recently made through strong lensing observations, which are becoming sensitive to substructures down to $10^{8} M_\odot$~\cite{vegetti_statistics_2009,vegetti_gravitational_2012,fadely_substructure_2012,hezaveh_detection_2016,bayer_observational_2018,gilman_warm_2019}. This is expected to improve even further in the near future with upcoming broad sky surveys~\cite{daylan_probing_2018,bechtol_dark_2019}. Similarly, recent astronomic observations opened novel opportunities for studying the small-scale distribution of DM within the Milky Way. These range from searches for gravitationally bound groups of metal-poor stars~\cite{belokurov_cats_2007,bechtol_eight_2015,simon_faintest_2019}, as was traditionally done to identify ultra-faint dwarf satellite galaxies, to more recently proposed approaches, such as analysis of gaps in stellar streams left behind tidally disrupted objects~\cite{erkal_properties_2015,banik_probing_2018}, searching for perturbations in the phase-space distribution of Milky Way's stars~\cite{feldmann_detecting_2015,buschmann_stellar_2018} or detecting tidal heating of gravitationally bound systems~\cite{penarrubia_stochastic_2019}.

The aim of this work is to explore a novel technique for possible detection of dark galactic subhaloes. Similarly to the approaches mentioned above, it relies on gravitational signatures of DM subhaloes imprinted in the spatial and kinematic distribution of surrounding stars. As it was already noted in~\cite{feldmann_detecting_2015}, subhaloes with masses $M_\s{sub} \gtrsim 10^8 M_\odot$ should induce observable features in the phase-space distribution of the galactic stellar disc and with the recent release of Gaia's astrometric data~\cite{lindegren_gaia_2018} it has become possible to search for them over a significant portion of our galactic neighbourhood. Furthermore, modern tools for reducing large and complex datasets opened new possibilities for analysing such extensive observations. For the problem at hand perhaps the most exciting progress comes from the field of computer vision where highly efficient convolutional neural network (CNN) models have been introduced, allowing for an incredibly sophisticated extraction of information from image-like data -- for a review on the topic see, e.g.,~\cite{ren_review_2019,rawat_deep_2017}.

In section~\ref{sec:survivability} we begin with a short review of various prediction for the subhalo abundance in Milky Way-like galaxies. In section~\ref{sec:stellar_field} we address the possibility of detecting signatures of dark subhaloes in the galactic stellar field using computer vision tools. First, we derive an analytical estimate for the detectability of perturbations induced by DM subhaloes. This is followed by a prescription for generating synthetic stellar fields, which are crucial for the training of neural networks, and finally, a concrete implementation of the CNN applicable to our problem is presented. Section~\ref{sec:results} contains several benchmarks of the advocated technique along with our main results. Finally, a summary of our findings and the prospects for future studies are presented in section~\ref{sec:conclusions}.

\section{Abundance of galactic subhalos}
\label{sec:survivability}

\begin{figure}
	\includegraphics[width=\columnwidth]{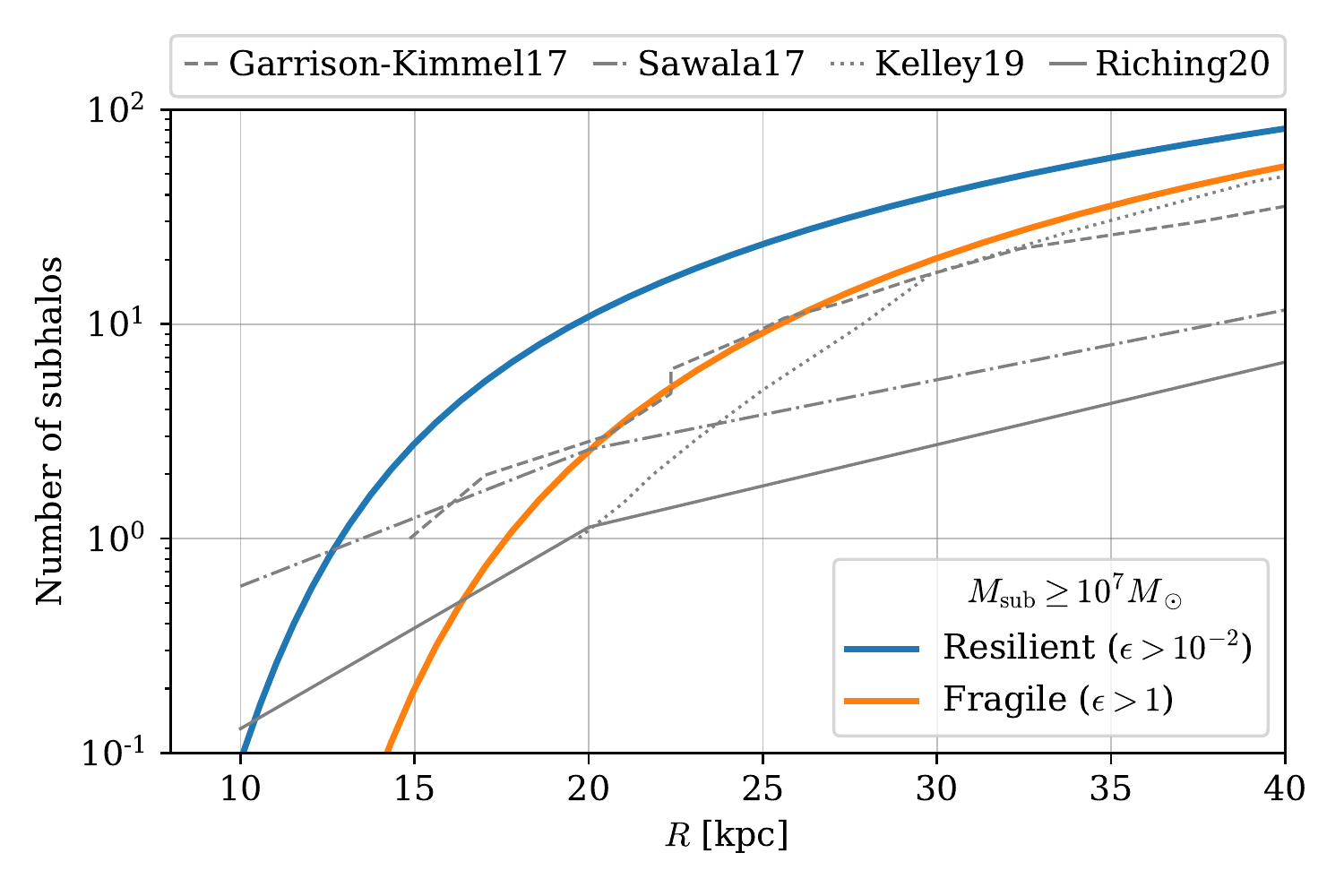}
	\centering
	\caption{The expected number of surviving cuspy DM subhaloes with $M_\s{sub} \geq 10^7 M_\odot$ in Milky Way-like galaxy as a function of galactocentric distance according to the semi-analytical model of \protect\cite{stref_remnants_2019}. The blue line corresponds to subhaloes which are highly resilient to tidal disruption effects (defined as bound until tidal radius reaches \protect\sfrac{1}{100} of their scale radius, i.e. $\epsilon > 10^{-2}$), while the orange line shows the results for haloes fragile subhaloes (considered disrupted when the tidal radius becomes equal to halo's scale radius, i.e. $\epsilon > 1$). For comparison we also show the cumulative number of subhaloes found in FIRE~\protect\cite{garrison-kimmel_not_2017} and APOSTLE~\protect\cite{sawala_shaken_2017,richings_subhalo_2020} hydrodynamic simulations, as well as the results of ELVIS dark matter only N-body simulations augmented by the gravitational potential of the baryonic disc~\protect\cite{kelley_phat_2019}.}
	\label{fig:subhaloes}
\end{figure}

Beside the microphysics of DM which determines the cosmological halo mass function, the number of galactic subhaloes also crucially depends on their resilience against tidal disruption and influence of baryons. These effects are expected to be particularly important in the inner part of our galaxy, which is, however, of the prime interest for our analysis as it hosts the majority of galactic stars. The abundance of DM subhalos with masses $M_\s{sub} \gtrsim 10^6 M_\odot$ was studied in several cosmological simulations -- see, e.g.,~\cite{yurin_stability_2015,zhu_baryonic_2016,garrison-kimmel_not_2017,sawala_shaken_2017,richings_subhalo_2020,kelley_phat_2019}. By comparing the results of DM-only simulations with the ones that include baryons (either full hydrodynamic simulations or N-body simulations augmented with the baryonic gravitational potential), they all conclude that the number DM subhalo is strongly suppressed within the inner 20 kpc of Milky Way-like galaxies. While~\cite{garrison-kimmel_not_2017,sawala_shaken_2017} found that there could exist a small number of surviving subhaloes with pericenters smaller then 20 kpc, their results were challenged by~\cite{kelley_phat_2019,richings_subhalo_2020} who found no surviving substructures with such small pericenters. On the other hand, dedicated high-resolution studies showed that remnants of cuspy DM subhaloes tend to survive the tidal forces caused by the gravitational potential of host halo and baryonic disc~\cite{penarrubia_effects_2005,errani_effect_2016, van_den_bosch_dark_2018,ogiya_dash:_2019,errani_can_2019}, even though they lose the majority of their mass during successive crossings of the pericenter. In particular,~\cite{van_den_bosch_dark_2018} showed that the disruption of subhaloes in cosmological simulations is often overestimated due to the resolution limits. Within the standard cold dark matter (CDM) paradigm an estimate for the number of surviving subhaloes within our galaxy can also be obtained using semi-analytic models~\cite{penarrubia_effects_2005,penarrubia_impact_2010,stref_modeling_2017,stref_remnants_2019}. According to them, several cuspy subhaloes with masses $M_\s{sub} \geq 10^7 M_\odot$ are expected to survive within the extent of the galactic stellar disc, regardless of the efficiency of tidal disruption.
Figure~\ref{fig:subhaloes} shows the expected number of subhaloes with $M_\s{sub} \gtrsim 10^7 M_\odot$ within a given galactocentric distance for the recent semi-analytic model of~\cite{stref_remnants_2019} as well as the counts obtained in several different cosmological simulations. In contrast with the cuspy subhalos, cored DM subhaloes are known to be disrupted much more easily and one does not expect to find them in the inner parts of the galaxies -- see, e.g.,~\cite{penarrubia_impact_2010,errani_effect_2016,errani_can_2019}. Consequently, the detection of dark (i.e. devoid of baryons) galactic subhaloes would be a strong indicator in favour of the CDM hypothesis, casting doubt on a wide range of alternative candidates, ranging from keV sterile neutrinos, ultra-light bosons to self-interacting DM models. On the other hand, the absence of the evidence for DM subhaloes with masses $M_\s{sub} \gtrsim 10^7 M_\odot$ would be a compelling argument against the CDM paradigm. However, in this case, robust conclusions would be much more difficult since not more then a few detectable CDM subhaloes are expected within the galactocentric distances that can be probed through the existing observations.

\section{A closer look at galactic stellar field}
\label{sec:stellar_field}


The Gaia mission has opened a new window for studying the properties of DM within our galaxy.
In this work we explore the possibility of inferring information about DM substructure through the analysis of a galactic stellar field within 10 kpc radius around the Sun where full 6D phase-space measurements are available for nearly 7 million stars~\cite{collaboration_gaia_2018}. Such approach was previously considered by~\cite{feldmann_detecting_2015,buschmann_stellar_2018}. While~\cite{feldmann_detecting_2015} studied the appearance, size and duration of the typical perturbations induced by a subhalo crossing the stellar disc using simulations, they did not suggest a concrete method for finding the signatures in actual observations. On the other hand,~\cite{buschmann_stellar_2018} proposed a statistical analysis for detecting the perturbations that is based on a comparison between the observations and an analytical phase-space distribution function for the stellar field in the vicinity of a subhalo. However, the latter can be obtained only for DM subhaloes with a cored Plummer density profile passing through a stellar field that is homogeneous and has isotropic Gaussian velocity distribution which are all very restricting assumptions.

In this work we present a novel approach of inferring the amount of dark substructures within the Milky Way that is applicable to any stellar distribution as well as an arbitrary DM density profile within the DM subhaloes.
In particular, we explore the ability of CNN to recognise the characteristic subhalo-induced patterns in the three-dimensional maps of stellar number density and moments of velocity distribution. As it is typically the case in all practical applications of machine learning, the success crucially depends on the quality of training data which has to be as representative of the real data as possible and at the same time provide a fair sampling of the configuration space that is of interest for the particular problem. Since the results presented here are only the first step in the exploration of prospects that CNNs offer in this context, only a simplified mock datasets will be addressed; we will neglect the perturbations of stellar field that might arise due to, e.g., globular clusters or spiral arms and consider only homogeneous stellar field with Gaussian velocity distribution. However, as discussed below, a generalisation of our approach to an arbitrary stellar phase-space distribution can be easily achieved.

In the following, we will first address the feasibility of detecting small perturbations in stellar kinematics sourced by DM substructures based on simple analytical estimate. This will be followed by a discussion of the algorithm for generating mock realisations of the galactic stellar field in the presence (or absence) of DM subhaloes, which will serve as training data for the CNN. Finally, the details regarding our implementation of a simple classifier network designed to detect the number of subhaloes in a given stellar field are presented.

\subsection{Detectability of subhaloes}
\label{sec:gaia_ml_intro}


There are many factors that determine whether the subtle perturbations in stellar phase-space distribution can be detected through given astrometric observations. Evidently, the mass of subhalo plays a crucial role since it is directly proportional to the force exerted on surrounding stars, however, also the relative velocity between the subhalo and the stellar field can be important. Additionally, stellar number density, kinematic properties and observational uncertainties also have to be taken into account since they determine the sample size and noise level for studying the perturbations.

To get a simple estimate for the magnitude of the gravitational perturbation caused by a DM subhalo, one can begin by computing the variation in stellar kinetic energy in the vicinity of a compact clump with mass $M_\s{sub}$:
\begin{align}
\label{eqn:perturbation}
\delta T_\s{sub} = \frac{G M_\s{sub}}{\Delta r} \;\;\;\;\;\; \textrm{with} \;\;\;\;\;\; \Delta r^{-1} = r_\s{min}^{-1} - r_\s{max}^{-1} \;,
\end{align}
where $r_\s{min}$ and $r_\s{max}$ are the minimal and maximal distance of the considered star to the centre of subhalo, while G is the gravitational constant. To detect the perturbation of size $\delta T$ the signal must be at least of the same order as the noise. Given a stellar population with velocity dispersion $\sigma_\star$ and typical measurement error $\delta v_\star$, one can estimate the average observational uncertainty in kinetic energy of a single star (assuming no net motion of stars):
\begin{align}
\label{eqn:obs_noise}
\delta T_\s{obs} \sim \sigma_\star \cdot \delta v_\star \, .
\end{align}
Since the velocities of individual stars are just random samples from their overall velocity distribution the observations become useful only when nearby stars are binned together and local variations in the stellar phase-space distribution are studied. This, however, induces additional shot noise due to the extraction of average quantities from a finite sample. For a Gaussian velocity distribution, one can show that the variance in the inferred velocity dispersion equals to:
\begin{align}
\label{eqn:shot_noise}
\mathrm{Var}(\sigma^2_\star) = \frac{2 \sigma_\star^4}{N_b - 1} \; ,
\end{align}
where $N_b$ is the number of stars in a given spatial bin. Therefore, by binning the stars in volume elements with characteristic length $\delta r$ and demanding that the signal in at least one bin is of the same order as the sum of shot noise and measurement errors, one obtains:
\begin{align}
\label{eqn:sensitivity}
\hat{M}_\s{sub} & \sim \frac{\sigma_\star \cdot \delta r}{G} \left( \delta v_\star + \sigma_\star \sqrt{\frac{2}{n_\star \cdot \delta r^3-1}} \right) \; ,
\end{align}
where we approximated $N_b \approx n_\star \cdot \delta r^3$ with $n_\star$ being the stellar number density. The above expression provides an order-of-magnitude estimate for the minimum subhalo mass, $\hat{M}_\s{sub}$, that could be detected in a given stellar field.
Note that in the above derivation we tacitly assumed that $\delta r$ is greater than the size of subhalo, which is not necessarily true, while for successful application of the suggested method $\delta r$ must be appreciably smaller than extent of the mapped stellar field. 
By using $\delta r$ that minimizes the expression~\eqref{eqn:sensitivity} for the typical values of $n_\star$, $\sigma_\star$ and $\delta v_\star$ inferred from the Gaia data one finds:
\begin{align}
\label{eqn:M_min_estimate}
\hat{M}_\s{sub} & \sim 10^7 \; M_\odot \cdot \left( \frac{\sigma_\star}{50 \; \s{km/s}} \right)^{5/3} \left( \frac{n_\star}{10^3 \s{kpc}^{-3}} \right)^{-1/3} \left( \frac{\delta v_\star}{5 \; \s{km/s}} \right)^{1/3} \; .
\end{align}
While the above estimate is obtained under favourable assumptions of point-like subhalo, negligible relative velocity between the subhalo and the stellar field, and signal-to-noise ratio of $\mathcal{O}(1)$, it non-the-less gives us reason to believe that sensitivities to phenomenologically interesting regime of $M_\s{sub}$ could be reached with high-quality astrometric measurements.
Furthermore, significant improvements in the coverage and accuracy of observations are expected with the future Gaia data releases~\cite{gilmore_gaia:_2018}, but also upon combining their measurements with other surveys such as the ones planned for LSST~\cite{drlica-wagner_probing_2019,bechtol_dark_2019} and other large telescopes~\cite{simon_testing_2019}.
At the same time, one can see from equation~\eqref{eqn:M_min_estimate} that the smallest detectable subhalo mass strongly depends on the $\sigma_\star$ and hence studying kinematically colder stellar populations, such as thin disc stars (or more accurately speaking, $\alpha$-poor population), could lead to significantly lower $\hat{M}_\s{sub}$. However, this comes at the cost of reducing the probed volume of the galaxy and hence lowering the probability to find a subhalo with $M_\s{sub} \gtrsim \hat{M}_\s{sub}$ in the considered region, while the associated decrease in the number of tracer stars with respect to the entire catalogue should have a sub-dominant effect due to the weak (inverse cubic root) dependence of $\hat{M}_\s{sub}$ on $n_\star$.

\subsection{Generation of mock data}
\label{sec:gaia_ml_mocks}

There are several possible approaches of producing mock stellar fields that contain a number of dark subhaloes with known properties. As already mentioned, analytical phase-space distribution functions can be computed, but only under very restricting assumptions which are inconsistent with realistic distribution of stars within the Milky Way. Another possibility is to perform numerical simulations and extract the stellar distribution and their response to subhaloes from them. However, this approach might be limited by the resolution of the simulations as well as turn out to be very computationally costly, especially if one wishes to produce a sufficient number of samples to train a neural network. For these reasons we will instead make use of an alternative approach in which the phase-space distribution of stars can be first adjusted to match the one inferred from observations and afterwards perturbed according to the gravitational effects of the embedded subhaloes.

\begin{figure}
	\centering
	\includegraphics[width=0.91\columnwidth]{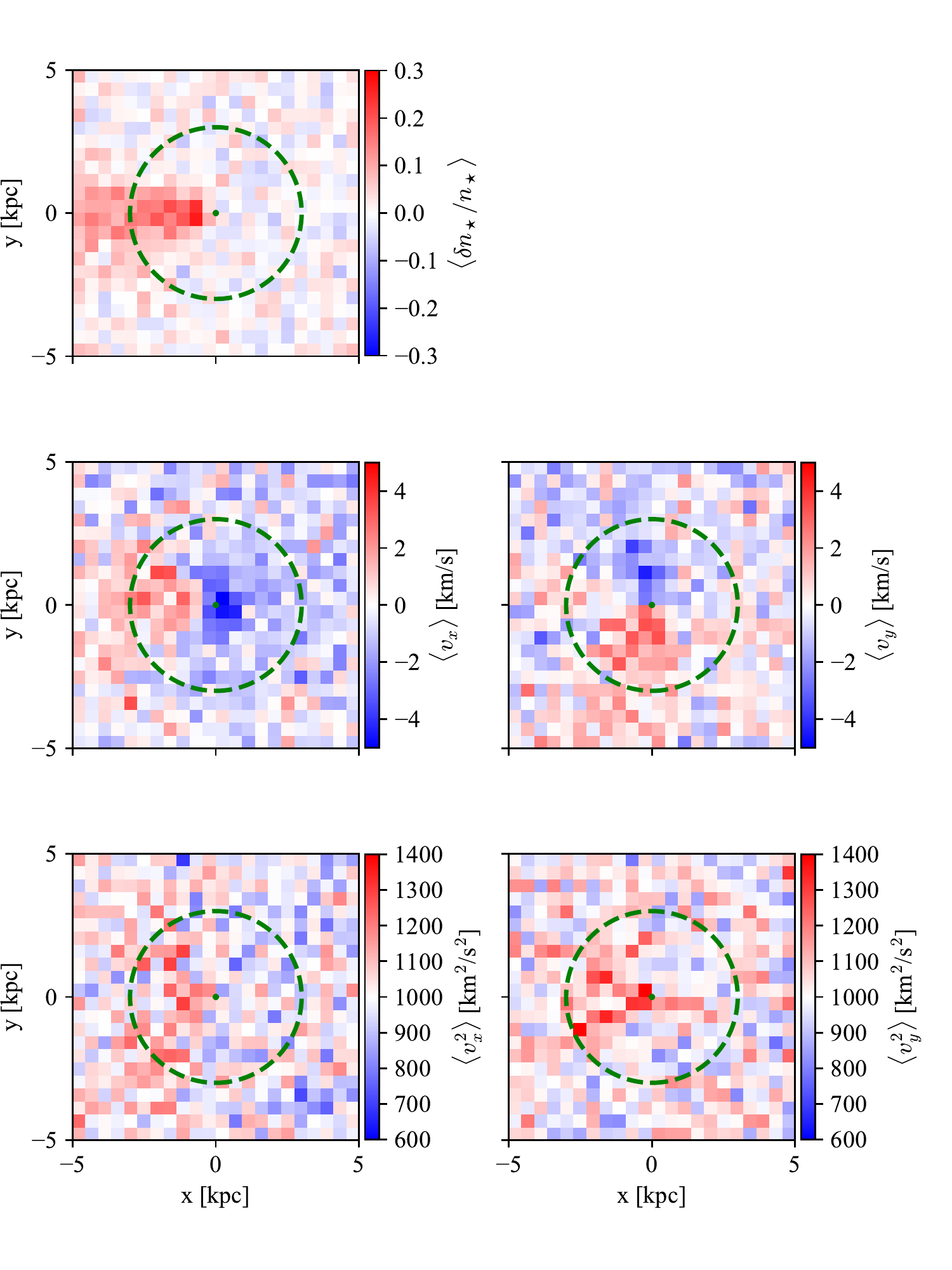}
	\includegraphics[width=0.91\columnwidth]{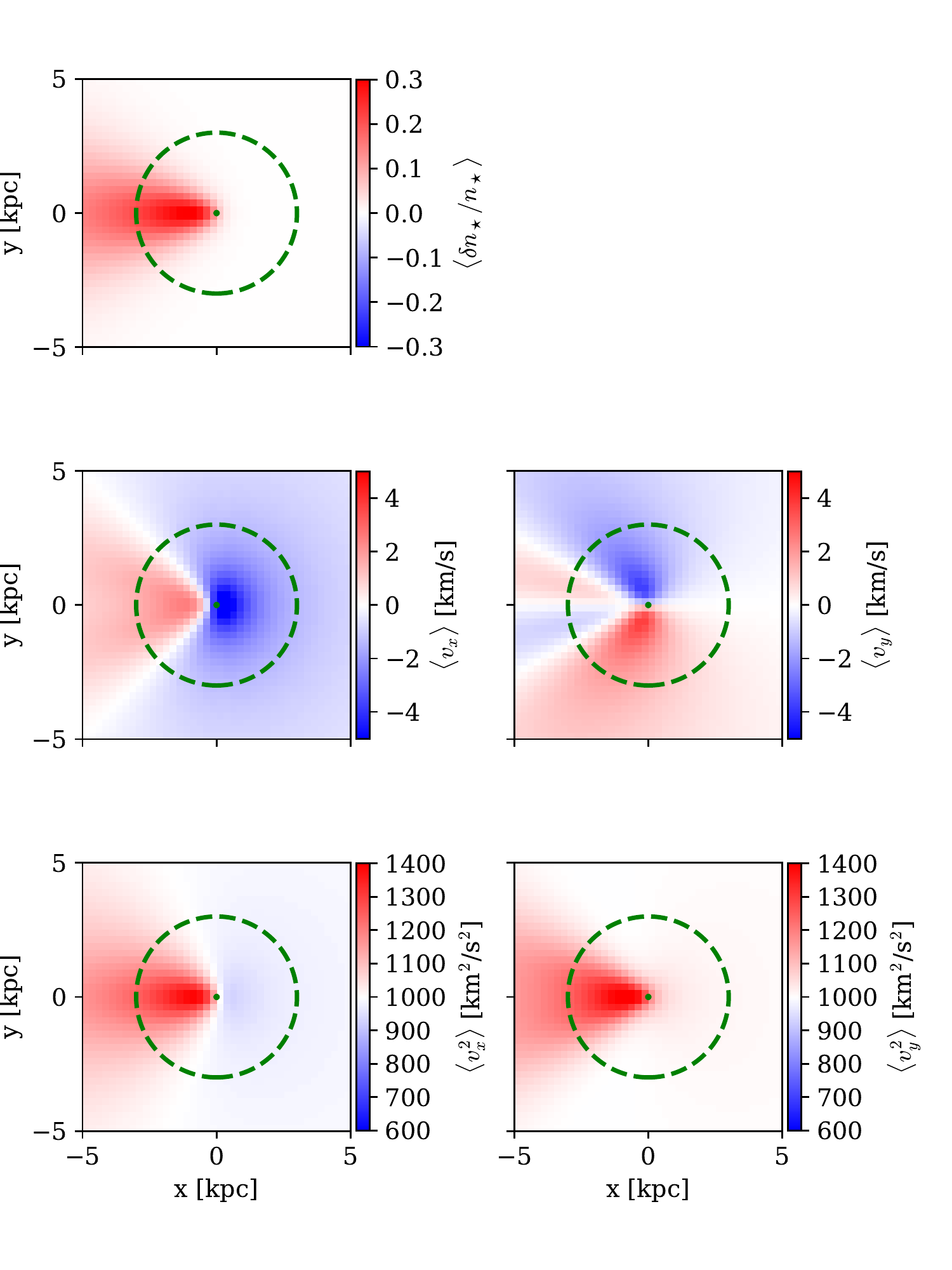}
	\caption{Binned stellar number density perturbation, $\langle \delta n_\star / n_\star \rangle$, average velocities, $\langle v_i \rangle$, and velocity dispersions $\langle v^2_i \rangle$ along parallel and perpendicular axis to the subhalo's movement direction. The top panel contains the results for stellar field generated according to the procedure described section~\ref{sec:gaia_ml_mocks} while the lower panel shows the results obtained from the analytic phase-space distribution function of \protect\cite{buschmann_stellar_2018}. In both cases the DM subhalo has a mass of $M_\s{sub} = 10^8 M_\odot$ and $|\vec{v}_\s{sub}| = 100$ km/s along $\hat{x}$ direction, while the fiducial stellar field has $n_\star = 10^4$ kpc$^{-3}$ and $\sigma_\star = 31.6$ km/s.}
	\label{fig:wakes}
\end{figure}

Such datasets can be constructed by first dividing the relevant volume of the galaxy in spatial bins, where each of them is populated with stars following the appropriate net properties, namely the local number density and velocity distribution. These can be inferred from the observations, however, they need to be averaged over scales much larger than the expected size of DM perturbation in order to avoid including unknown structures in the synthetic stellar field. In practice, this can be achieved by binning the mapped stars in relatively large spatial bins and then smoothly interpolating the number density and moments of velocity distribution over a finer grid that is used for constructing the mock dataset. Such a smoothed stellar field then needs to be endowed by DM substructure and its phase-space distribution updated according to the gravitational effects of the inserted subhaloes. To achieve this one first needs to specify the subhalo position, $\vec{r}_\s{sub}$, velocity, $\vec{v}_\s{sub}$, and mass, $M_\s{sub}$. The latter allows one to determine the characteristic radius, $r_g$, at which the gravitational potential of the subhalo is comparable to the typical uncertainties in stellar kinematics. Making use of similar reasoning as in derivation of the relation~\eqref{eqn:M_min_estimate} one can define:
\begin{align}
r_g \equiv \frac{G M_\s{sub}}{\sigma_\star \left( \delta v_\star + \sigma_\star \sqrt{\frac{2}{N_b-1}} \right)} \, ,
\end{align}
which is to be used for determining whether a given star at position $\vec{r}_\star$ is considered as perturbed or the gravitational effects of the subhalo can be safely neglected. While such criterion guarantees that the change in stellar kinetic energy is below the typical measurement error, one could worry that there is a non-negligible effect on the direction of its velocity, if the star passed sufficiently close to the subhalo in the past. However, the probability for such a close encounter falls of as $(\vec{r}_\star - \vec{r}_\s{sub})^{-2}$ and can also be safely neglected. On the other hand, stars that are within the characteristic radius, i.e. $\left| \vec{r}_\star - \vec{r}_\s{sub} \right| < r_g$, receive non-negligible correction, which can be computed according to the standard results for scattering in a central potential.
This procedure of synthesizing the Milky Way's stellar field is robust against inhomogeneities in fiducial distribution, provided that they correspond to scales sufficiently larger than $r_g$. Furthermore, it is also capable of accounting for spatial dependence of the stellar velocity distribution, which does not need to be Gaussian nor isotropic. 
However, in this preliminary study we will consider only the simplified case of homogeneous stellar field with Gaussian velocity distribution. An example of the produced dataset is presented in figure~\ref{fig:wakes}, along with the corresponding results obtained from the analytical model. As can be seen from the plots, qualitative features in the obtained stellar density and moments of the velocity distribution are very similar, despite the fact that significantly different methods were used to generate the maps. Clearly, the key difference is that the maps generated through our procedure contain shot noise due to finite sample of stars, as it is the case also in real observations, while the analytical method can be evaluated with arbitrary precision.

\subsubsection{Limitations of the outlined approach}

Addressing the actual stellar field, as mapped by the Gaia satellite, is a complicated task. Beside spatial variations in the distribution of stars, the data contains numerous features which are most probably not sourced by DM substructures, such as globular clusters, over-densities related to giant molecular clouds and spiral arms or dynamical structures induced by the galactic bulge. 
Inclusion of these features would require much more careful generation of training data, but also make the extraction of DM signatures more difficult. On the other hand, there are several reasons why it should be possible to discriminate the signatures induced by DM subhaloes from other perturbations. For example, dynamical structures should exhibit appreciably different and less localised features in the stellar phase-space distribution compared to the perturbations induced by compact spherical DM subhaloes. However, to obtain quantitative results, further improvements in the modelling of galactic stellar field and perhaps a comparison with dedicated simulations will be needed. Similarly, false detections due to the presence of globular clusters can be avoided since they are composed of stars that are kinematically colder than the surrounding stellar field which should partially mask the effect of gravitational heating exerted on the nearby stars. Furthermore, it should also be possible to distinguish globular clusters from DM subhaloes in dedicated follow-up analysis of the observations. At the same time globular clusters, as well as giant molecular clouds, rarely exceed masses of $10^6 M_\odot$ (for an up-to-date catalogue of globular clusters see, e.g.,~\cite{hilker_galactic_2019}) and would, therefore, produce features that are too faint to be detected by our method with the currently available data. Another possible concern is the presence of multiple subhaloes with overlapping regions of influence, which also can not be accurately modelled within the proposed prescription of generating training datasets. However, such clumping of subhaloes seems unlikely to survive the effect of tidal forces in the inner part of the galaxy. In case such systems of multiple subhaloes would survive the tidal effects, they would have to form very compact clusters, and hence their effect on surrounding stars would be similar to that of a single DM subhalo with somewhat lower total mass (i.e. presence of additional substructure would reflect in less peaked gravitational potential which would induce smaller perturbations in the bin-averaged properties of the stellar filed).

\subsection{Search for DM subhalo signatures using CNN}
\label{sec:gaia_ml_results}

Identifying gravitational perturbations induced by DM subhaloes through the means of traditional statistical tools turns out to be challenging even in a simplified setting. 
Therefore, we in this work resort to CNNs which are particularly powerful tool for extracting characteristic features in image-like data. They also exhibit the property of transnational invariance which allows them to detect the sought-after patterns regardless of their spatial position within the input image. Furthermore, the machine learning algorithms use training data to teach themselves what are the relevant features and the user is only required to prepare representative training datasets, which is in the case of galactic stellar field much simpler than constructing accurate parametric models. As an additional benefit, the machine learning approach is typically much less computationally demanding than careful sampling of the likelihood function and many highly optimised implementations of deep neural networks have been recently developed and made publicly available.

The CNN that we used in this study takes the moments of stellar phase-space distribution as its input and returns the probabilities that a certain number of subhaloes are present in a given stellar field. Since no more than a few subhaloes with mass greater than $\hat{M}_\s{sub}$ are expected within the portion of galaxy mapped by Gaia~\footnote{From figure~\ref{fig:subhaloes} one can see that less then 10 subhaloes with $M_\s{sub} \gtrsim 10^7 M_\odot$ are expected within the inner 20 kpc of the Milky Way. However, since the stellar field mapped by Gaia covers only a small fraction of this volume we find it sufficient to consider only 3 output classes for the network, which correspond to realizations with 0, 1 and 2 embedded subhaloes. On the other hand, we explicitly checked that increasing the number of classes has a negligible effect on the performance of the network.}, we use a classifier network with only a three output classes, where each of them corresponds to a fixed number of subhaloes contained in the given volume. On the other hand, the inputs comprise of seven 3D grids which correspond to spatially binned information regarding the stellar number density, as well as average velocities and velocity dispersions along the three coordinate axes. These seven grids are referred to as channels since in the conventional application of CNNs they usually represent different colour channels of the image. By construction CNNs cross-correlate the signals at a given spatial position across all the channels which in the case of our application results in high sensitivity for the localized perturbations induced by the DM subhaloes. Furthermore, the translational invariance of CNNs greatly reduces the number of required training samples, since the spatial position of the signal become irrelevant, and the network is automatically capable of recognising it at any location within the input grid. Beside just extracting the number of dark subhaloes in a given dataset, it would also be possible to transform the described neural network from a classifier into a regression model capable of quantifying the mass of perturber, as well as its location and velocity. However, in this preliminary study, we will consider only the simpler case of classification network, since regression models can be constructed only for a known fixed number of embedded subhaloes and would, therefore, serve only as second step in the analysis of the observed stellar field.

The network used in this study was constructed through Python implementation of the Keras library~\cite{chollet_et_al._keras_2015}. In our tests we found the optimal balance between the network's complexity and performance when using two 3D convolutional layers, each followed by batch normalisation, drop-out and max-pooling layer, that in the end connected to two fully connected (i.e. dense) layers. Each convolutional layer consists of a number of filters that are trained to produce positive signals when convoluted with a portion of the grid that contains the sought-after patterns. For our task we found it sufficient to use 32 filters, which performed best when chosen to be $3 \times 3 \times 3$ cubes. The outputs of the convolution were then fed to the activation function, for which we used the rectified linear unit (ReLU) function $f(x) = \s{max}(0, x)$, producing the final output of the convolutional layer. To improve the training efficiency of the network, each convolutional layer was followed by a normalisation layer, ensuring that the mean activation of neurons is close to 0 with a standard deviation of 1. The normalised outputs were subsequently fed to the drop-out layer, which randomly disables a given fraction of neurons at each training epoch to increase robustness of the network as well as improve the convergence of the training procedure. We found the best performance by setting the disabled fraction of neurons to 25\%. Finally, the drop-out layers were followed by max-pooling layers which return the maximum value of each $2 \times 2 \times 2$ sub-grid of the input. This is crucial for reducing the amount of data passing through the network and ensuring that one ends up with a manageable number of weights in the final layers. In the last part of the network the outputs of second convolutional block were fed to a dense layer, where each input neuron is connected to all the output neurons with weights that are appropriately adjusted during the training procedure. The role of the dense layers is to perform the final classification of the inputs and for our purpose we found that two of them with 64 and 32 nodes were sufficient. A precise summary of the network layout is presented in table~\ref{tab:nn_layout}. For training of the network Keras implementation ADAM optimiser with learning rate $\texttt{lr} = 10^{-4}$ was used, along with the categorical cross-entropy loss function. Throughout this work each dataset consisted of 32000 realisations of stellar field and was split into three parts, namely \sfrac{3}{4} of the samples were used as a training set, \sfrac{1}{8} as a validation set and the remaining \sfrac{1}{8} as a test set. The validation set was used to benchmark the networks performance during the training procedure and when the loss function on it started increasing the training was stopped to avoid over-fitting. Finally, the test set was used to benchmark the performance of network after the training was completed.

\begin{table}
	\centering
	\begin{tabular}{c|c|c}
		Layer & Output shape & Number of parameters \\ 
		\hline 
		Conv3D & (7,20,20,20,32) & 6080 \\ 
		BatchNormalization & (7,20,20,20,32) & 80 \\ 
		Dropout & (7,20,20,20,32) & 0 \\
		MaxPooling & (7,10,10,10,32) & 0 \\ 
		Conv3D & (7,8,8,8,32) & 27680 \\ 
		BatchNormalization & (7,4,4,4,32) & 16 \\ 
		Dropout & (7,20,20,20,32) & 0 \\
		MaxPooling & (7,4,4,4,32) & 0 \\
		Flatten & 14336 & 0 \\ 
		Dense & 64 & 131136 \\ 
		Dense & 32 & 2080 \\
		Dense & 3 & 99 \\
	\end{tabular}
	\caption{Summary of the CNN that was used in this work. In total the model has 167091 trainable parameters.}
	\label{tab:nn_layout}
\end{table}

\section{Results}
\label{sec:results}

In the following we present our findings regarding the performance of CNNs for detecting perturbations in the galactic stellar field sourced by traversing DM subhaloes. First we address the particularly simple case of a homogeneous and isotropic stellar field containing slow-moving point-like subhaloes to establish the optimal sensitivity that could be reached by this method and compare it to the analytical estimate derived in section~\ref{sec:gaia_ml_intro}. Subsequently, we move to a more realistic setup, adopting a simulation-motivated velocity distribution for the subhaloes as well as treat them as extended objects. To make our tests as realistic as possible, we adopt the typical number density and kinematic properties of stars as mapped by the Gaia satellite and consider theoretically motivated mass-concentration distribution of subhaloes.

\subsection{Simplified setup}

\begin{figure}
	\includegraphics[width=\columnwidth]{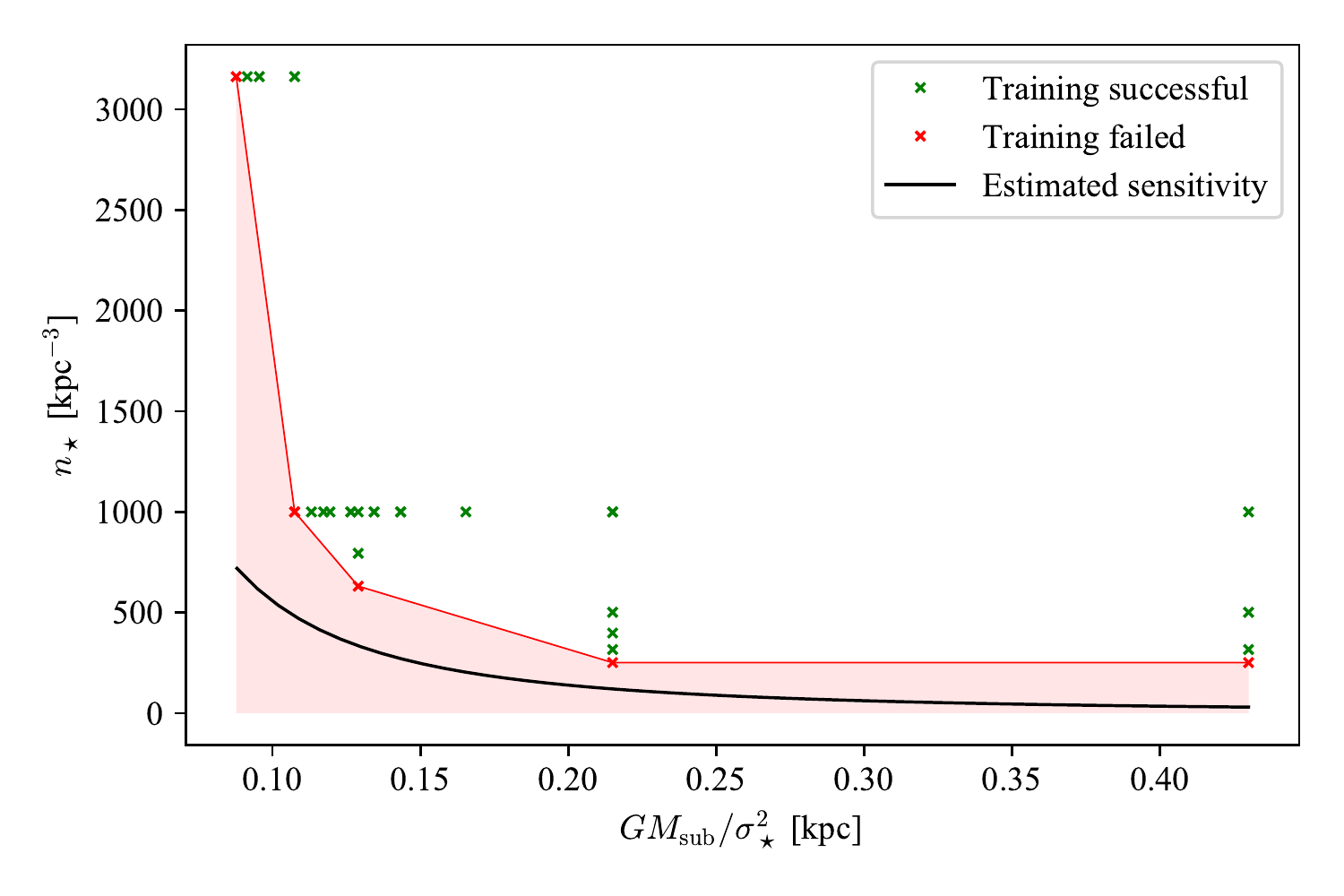}
	\caption{The ability of CNN to successfully train at given point in the parameters pace spanned by subhalo mass over stellar velocity dispersion, $G M_\s{sub} / \sigma^2_\star$, and stellar number density, $n_\star$. Green crosses mark the points where network successfully trained and reach accuracy above 95\%, while red crosses mark the points where network failed to train, resulting in accuracy $\sim 33$\%. The red region marks the part of parameter space, where the CNN approach is expected to fail, while the black line denotes the sensitivity threshold estimate from equation~\eqref{eqn:sensitivity}, assuming $\delta v_\star = 0$.}
	\label{fig:sensitivity}
\end{figure}

As already stated above, the scope of this preliminary study was limited to a homogenous stellar field with Gaussian velocity distribution and slow moving point-like subhaloes ($\bar{v}_\s{sub} \lesssim \sigma_\star$) in order to compare the methods efficiency with the theoretical limit~\eqref{eqn:sensitivity}. In such setting, each dataset can be fully characterised by the number of embedded subhaloes and their masses, and additionally the stellar number density and velocity dispersion.
To explore the dependence of the detection efficiency on $M_\s{sub}$, but also on stellar parameters, we trained CNNs for several combinations of $M_\s{sub} / \sigma_\star^2$ and $n_\star$. Results are shown in figure~\ref{fig:sensitivity}, where the green crosses mark the points in the parameter space where the network trained successfully, reaching accuracy~\footnote{The accuracy used to benchmark the performance of the network throughout this work is defined as the percentage of test stellar fields where the network gave the highest probability to the class with the correct subhalo count.}  above 95\%, and the red crosses where it failed to reach accuracy better than $\sim 33$\% which is equivalent to random guessing. In the same plot we also show the sensitivity estimate from equation~\eqref{eqn:sensitivity} with $\delta v_\star$ set to 0 since the mock datasets did not include measurement uncertainties of the stellar velocities~\footnote{It turns out that the errors in velocity measurements provided by Gaia data are subdominant and the shot noise due to binning dominates the uncertainty in inferred moments of stellar velocity distribution. An exception to this is perhaps the stellar field within 1 kpc distance from Earth, where the number density of mapped stars is much larger.}. Unsurprisingly, the analytical estimate predicts better sensitivity than the one found in practice which implies that the perturbations in stellar distribution need to be somewhat larger than the shot noise induced by the binning of the stars in order for the CNN to detect them. At intermediate values of $G M_\s{sub} / \sigma^2_\star \sim 0.2$ kpc, for which the $\delta r$ that was used for binning the stellar field is near the optimal choice, a correction factor in $\hat{M}_\s{sub}$ of less than 2 is needed. However, the discrepancy gets more sever at larger and smaller values of $G M_\s{sub} / \sigma^2_\star$. This could be perhaps mitigated by choosing more appropriate $\delta r$, but that would also require either changing the extent of the analysed stellar field or modifying the size of network layers due do a different number of spatial bins.

Additionally, we also explored the case of anisotropic stellar velocity distribution by setting different values of velocity dispersion along the three spatial directions. In this case, we found that the sensitivity mostly depends on the smallest velocity dispersion component (denoted as $\sigma_{\star,z}$), but gradually decreases as the velocity dispersion in the perpendicular components (denoted as $\sigma_{\star,\s{xy}}$) increases. The maximal increase in $\sigma_{\star,\s{xy}}$ that still lead to classification accuracy better then 95\% was found to depend non-trivially on $M_\s{sub}$, $\sigma_{\star,\s{z}}$ and $n_\star$, however, datasets with $\sigma_{\star,xy} / \sigma_{\star,z} \lesssim \sqrt{2}$ typically resulted in only few percent lower accuracy then the isotropic datasets with $\sigma_\star = \sigma_{\star,\s{z}}$.

\subsection{A step closer towards the Milky Way}

\begin{figure}
	\centering
	\includegraphics[width=0.8\columnwidth]{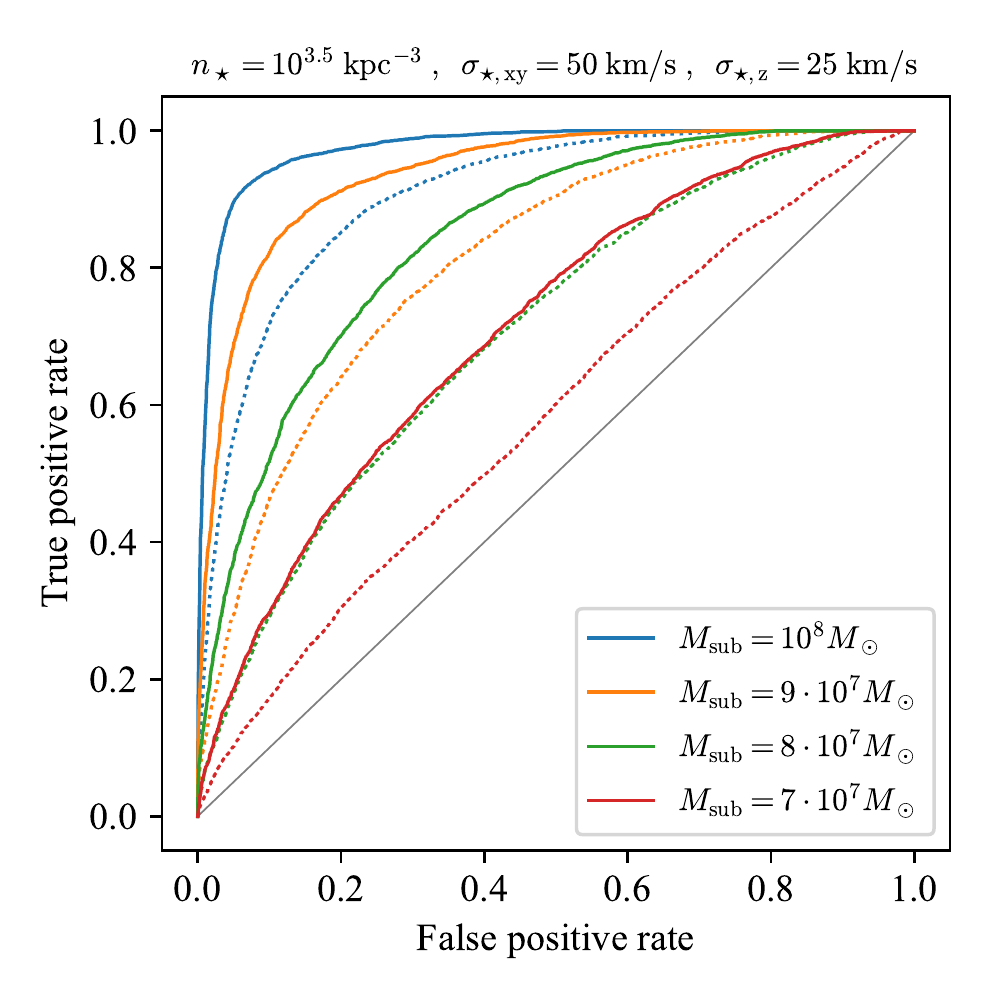}
	\caption{ROC curves for CNNs trained on datasets with different subhalo masses but same fiducial stellar field parameters, namely $n_\star = 10^{3.5}$ kpc$^{-3}$, $\sigma_{\star,xy} = 50$ km/s and $\sigma_{\star,z} = 25$ km/s. The full lines correspond to point-like subhaloes while the dashed lines were obtained for subhaloes with NFW density profile truncated at $r_s$.}
	\label{fig:roc_mw1}
\end{figure}

To further explore the performance of our method we subsequently applied it to mock stellar fields with realistic kinematic properties and refined modelling of DM substructures. The first thing to note is that DM subhaloes in the inner part of our galaxy are expected to have relatively high velocities which makes their detection more difficult compared to the tests presented above. To account for this, we adopted the assumption that their velocities follow a Gaussian distribution with the mean $\bar{v}_\s{sub} = 300$ km/s and standard deviation of $\sigma_\s{sub} = 80$ km/s, which is similar to the velocity distributions found in hydrodynamic simulations at galactocentric distances between 10 and 20 kpc~\cite{sawala_shaken_2017,richings_subhalo_2020}. While the simulations also suggest that their velocity distribution is anisotropic with a notable preference for tangential orbits, we believe that this has a much smaller effect on the sensitivity of our method and can be postponed to future studies where also more sophisticated modelling of the stellar field anisotropies will be used. The second improvement in the modelling comes from the fact that realistic DM subhaloes are extended objects which generically leads to fainter signatures with respect to the point-mass approximation that has been considered so far. Therefore, we examined the sensitivity of our method to subhaloes that follow the simulation motivated NFW density profile~\cite{navarro_universal_1997} with initial concentration parameter given by the mean value of the mass-concentration relationship of~\cite{sanchez-conde_flattening_2014}. We further assumed that the subhaloes are stripped by the tidal forces down to the scale radius, i.e. $r_\s{t} / r_\s{s} = 1$, while their internal structure was assumed to remain unchanged. The particular choice of $r_\s{t}$ was motivated by the fact that picking smaller values and retaining same $M_\s{sub}$ would result in objects closer to the point-mass approximation, while subhaloes with $r_\s{t} > r_\s{s}$ are expected to be increasingly rare since their concentration should be far greater than the mean value in the relevant mass range.
Regarding the stellar field, we used number density corresponding to the lower bound of Gaia's coverage along the galactic disc, $n_\star = 10^{3.5}$ kpc$^{-3}$, and average kinematic properties of the observed stars, namely $\sigma_{\star,\s{xy}} = 50$ km/s along the galactic plane and $\sigma_{\star,\s{z}} = 25$ km/s in the direction perpendicular to the galactic plane. In figure~\ref{fig:roc_mw1} we show the resulting ROC curves~\footnote{The efficiency of classifier networks can be nicely captured in receiver operating characteristic (ROC) curves, measuring the rate of correct classification versus the rate of false classifications as the threshold value for assigning the input to a certain category is being varied.} obtained for point-mass (solid lines) and extended (dashed line) DM subhaloes of various masses. As expected, the resulting sensitivities are worse than in the previously studied idealised cases.
The first reason for this is that even for subhaloes with masses above the sensitivity threshold some of the objects escape detection due to their high relative velocity with respect to the stellar field. Secondly, as can be seen from the plot, the performance of the network further decreases if one considers extended DM subhaloes which is a consequence of shallower gravitational potential in comparison with the point-mass case.
In this more realistic setup, CNNs reached the accuracy of only 76\% for $10^8 M_\odot$ subhaloes with extended mass profile, while the point-like objects of same mass perform somewhat better, reaching 90\% accuracy.

\begin{figure}
	\centering
	\includegraphics[width=0.8\columnwidth]{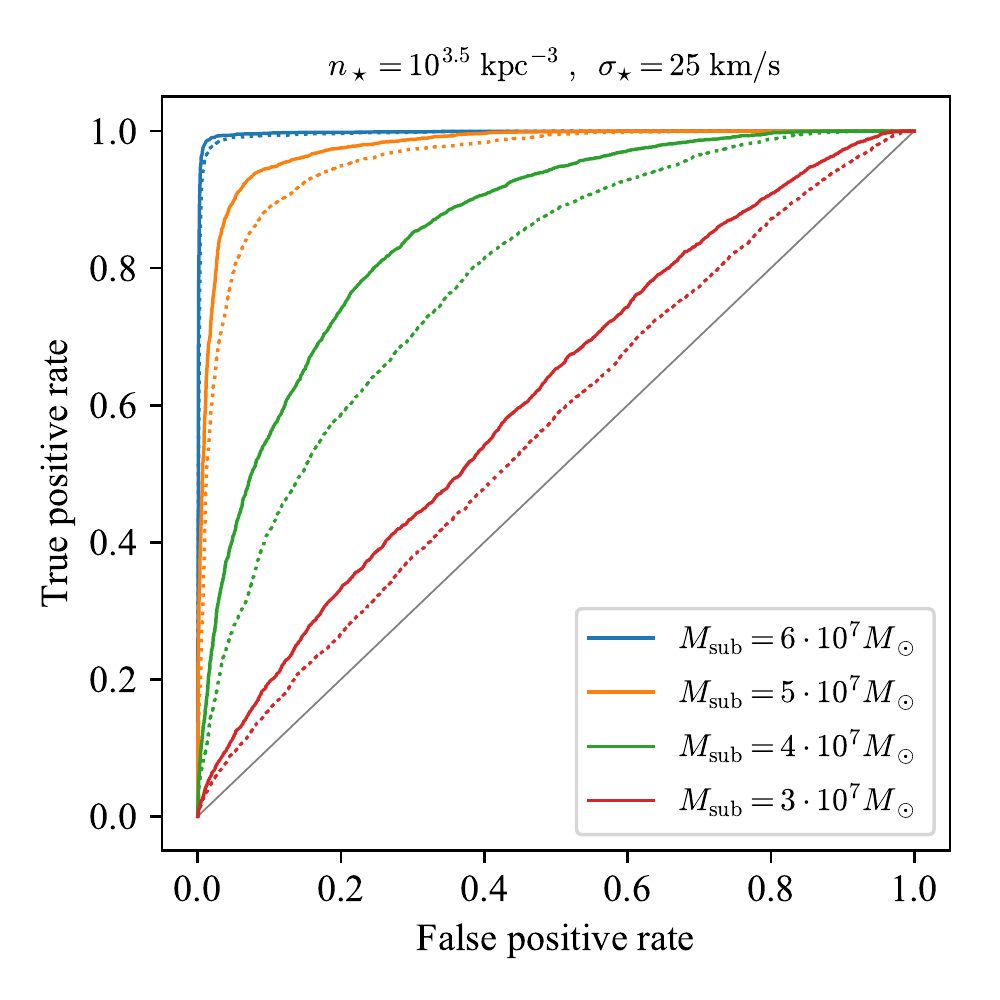}
	\caption{ROC curves for CNNs trained on datasets with different subhalo masses but same fiducial stellar field parameters, namely $n_\star = 10^{3.5}$ kpc$^{-3}$ and $\sigma_\star = 25$ km/s. The full lines correspond to point-like subhaloes while the dashed lines were obtained for subhaloes with NFW density profile truncated at $r_s$.}
	\label{fig:roc_mw2}
\end{figure}

To highlight the benefits of analysing cold stellar populations we also consider setup with $\sigma_\star = 25$ km/s, corresponding to approximate value of the velocity dispersion found for the thin disc stars -- see, e.g.,~\cite{rowell_white_2011,torres_random_2019}. The resulting ROC curves are shown in figure~\ref{fig:roc_mw2}. As can be seen from the plot, in this case the performance of the CNN is significantly better, reaching 85\% accuracy for extended DM subhaloes with $M_\s{sub} = 5 \cdot 10^7 M_\odot$ and 90\% accuracy for point-mass objects with the same mass. However, by restricting ourself to only a subpopulation of the mapped stars, one ends up with a lower number density of the stellar field, which naturally leads to worse performance of our method. To address this possible issue we also tested stellar fields with a roughly three times lower number density of stars, namely $n_\star = 10^3$ kpc$^{-3}$. In this case we found that the network reaches 85\% accuracy in identifying NFW subhaloes with $M_\s{sub} = 9 \cdot 10^7 M_\odot$. This loss of accuracy is somewhat worse than what would be expected from the scaling in equation~\eqref{eqn:sensitivity} and most probably suggests that larger spatial bins should have been used in the analysis. However, since the majority of galactic stars belong to the thin disc population, the decrease in sensitivity due to focusing only on this subset of stars would be much milder. This clearly shows that using the thin disc population of stars is preferred as it allows one to explore more phenomenologically interesting regime of subhalo masses.
In order to make our study even more realistic, we also considered the case where subhalo masses and concentration parameters were drawn from theoretically motivated probability distribution function. By limiting our attention to the mass range $M_\s{sub} \in [4 \cdot 10^7, 10^8] M_\odot$, where the network is known to perform reasonably well for monochromatic subhalo masses (i.e. identical $M_\s{sub}$ for all subhaloes), we assumed that the initial halo mass function follows a power-law with spectral index of $\alpha = -1.9$, corresponding to the model of~\cite{sheth_large-scale_1999}, and that the halo concentrations are distributed according to the log-normal mass-concentration relationship reported by~\cite{sanchez-conde_flattening_2014}. To account for the tidal stripping of the objects, we adopted the appropriate truncation radii that were provided by~\cite{stref_remnants_2019} through private correspondence. The resulting network reached an accuracy of 87\%, giving further support for the method's applicability to dark subhalo searches within the Milky Way. The achieved accuracies for detecting extended subhaloes of various masses in different stellar environments are summarised in table~\ref{tab:results}.

\begin{table}
	\begin{tabular}{c|c|c|c|c}
		$M_\s{sub}$ [$M_\odot$] & $\sigma_{\star,\s{xy}}$ [km/s] & $\sigma_{\star,\s{z}}$ [km/s] & $n_\star$ [kpc$^{-3}$] & accuracy \\ \hline \hline
		$10^8$ & 50 & 25 & $10^{3.5}$ & 76\% \\
		$9 \cdot 10^7$ & 50 & 25 & $10^{3.5}$ & 62\% \\
		$8 \cdot 10^7$ & 50 & 25 & $10^{3.5}$ & 50\% \\ \hline
		$6 \cdot 10^7$ & 25 & 25 & $10^{3.5}$ & 97\% \\
		$5 \cdot 10^7$ & 25 & 25 & $10^{3.5}$ & 85\% \\
		$4 \cdot 10^7$ & 25 & 25 & $10^{3.5}$ & 58\% \\ \hline
		$9 \cdot 10^7$ & 25 & 25 & $10^{3}$ & 85\% \\
		$8 \cdot 10^7$ & 25 & 25 & $10^{3}$ & 71\% \\
		$7 \cdot 10^7$ & 25 & 25 & $10^{3}$ & 56\% \\ \hline
		$4 \cdot 10^7$ -- $10^8$ & 25 & 25 & $10^{3.5}$ & 87\%
	\end{tabular}
	\caption{The resulting accuracy of the network for various values of subhalo masses (assuming truncated NFW density profile) and stellar field parameters.}
	\label{tab:results}
\end{table}

\section{Conclusions}
\label{sec:conclusions}

The amount of DM substructure within galaxies could provide crucial new information regarding the nature of DM particles. In this work, we present a novel method for constraining the abundance of subhaloes within the Milky Way through machine-learning-based analysis of the galactic stellar field. Astrometric data of adequate quality for such studies has only recently become available, but increasingly extensive and accurate observational catalogues are expected to be released in the coming years. 

After briefly reviewing various predictions for the subhalo abundance, we first derive a theoretical estimate for the optimal sensitivity of the outlined approach. We continue by providing an efficient and flexible method for generating stochastic realisations of stellar fields with arbitrary large-scale properties that are perturbed by traversing DM haloes. We then use this method to build the datasets required for training and testing the performance of CNNs in detecting the signatures of subhaloes. To begin with, we established the sensitivity that can be expected in a particularly favourable setting, namely homogenous and isotropic stellar field perturbed by slow-moving point-mass objects. As can be seen from figure~\ref{fig:sensitivity}, our results show that in this case the performance of the neural network is not far from the estimated ideal sensitivity given by expression~\eqref{eqn:sensitivity}. Subsequently, we turned our attention to more realistic setups with simulation-motivated velocity distribution and density profiles of DM subhaloes. Both of these features lead to fainter perturbations in the stellar field which consequently increase the threshold mass for successful detection. By adopting typical kinematic properties of Milky Way's disc stars, we found accuracy greater than 76\% only for identifying DM subhaloes with $M_\s{sub} \geq 10^8 M_\odot$. On the other hand, in agreement with the analytical estimate~\eqref{eqn:sensitivity}, our results show much better prospects for detecting low subhalo masses by focusing on cold stellar populations. Most likely the prime candidate are the galactic thin disc stars (or more precisely $\alpha$-poor population) since they are characterized by particularly low velocity dispersion, for which accuracy better then 85\% was obtained down to $M_\s{sub} \sim 5 \cdot 10^7 M_\odot$ given the same stellar number density, $n_\star$, and $M_\s{sub} \sim 9 \cdot 10^7 M_\odot$ assuming three times lower $n_\star$. Our results also show that the network is capable of reaching nearly the same accuracy when applied to stellar fields containing subhaloes with monochromatic and continuous mass distribution, as long as their mass is above the detection threshold. On the other hand, before being able to analyse actual observations, improvements are still needed in the generation of mock stellar catalogues to include all possible features which are not sourced by DM substructure. Nuisance signals might arise, for example, from globular clusters, giant molecular clouds, spiral arms or other galactic structures. While we list a number of arguments why it should be possible for CNNs to distinguish them from subhalo-induced phase-space patterns, this still needs to be confirmed through further tests.

There are also several possible improvements to our analysis. For example, better sensitivity could be perhaps achieved by considering a projection of the stellar motion on the sky sphere as it would allow one to use the entire Gaia dataset, without restricting only to a subset of stars with determined radial motions. However, it is not clear that this approach would necessarily result in better accuracy because stars with only 5D astrometric data carry less information and are typically associated much larger observational uncertainties in the measured parallaxes and proper motions. Further improvements might also be achieved by deploying different neural network models. For example, one could try using much deeper residual networks~\cite{zhang_deep_2015} or other types of neural networks. In future work we also plan to explore the possibilities of applying regression CNNs for extracting the mass, location and velocity of a given subhalo.

\section*{Acknowledgments}

We thank Gabrijela Zaharijas and Christopher Eckner for many stimulating debates and to Martin Stref and Ga\'etan Facchinetti for sharing their results regarding the survivability of DM subhaloes within Milky Way.
The author acknowledges partial support from the European Union's Horizon 2020 research and innovation programme under the Marie Sk\l odowska-Curie grant agreements No 690575 and 674896, as well as the ANR project ANR-18-CE31-0006.




\bibliographystyle{mnras}
\bibliography{citations} 








\bsp	
\label{lastpage}
\end{document}